\documentclass
[showpacs,pra,onecolumn,12pt,titlepage,superscriptaddress]{revtex4}%
\usepackage{amsfonts}
\usepackage{amsmath}
\usepackage{amssymb}
\usepackage{graphicx}%
\providecommand{\U}[1]{\protect\rule{.1in}{.1in}}
\newtheorem{theorem}{Theorem}

\newtheorem{corollary}[theorem]{Corollary}

\newenvironment{proof}[1][Proof]{\noindent\textbf{#1.} }{\ \rule{0.5em}{0.5em}}
\begin{document}
\title{Local Distinguishability of Orthogonal Quantum States and Generators of SU(N)}
\author{Ming-Yong Ye}
\affiliation{Key Laboratory of Quantum Information, Department of Physics, University of
Science and Technology of China, Hefei 230026, People's Republic of China}
\affiliation{School of Physics and Optoelectronics Technology, Fujian Normal University,
Fuzhou 350007, People's Republic of China}
\author{Wei Jiang}
\affiliation{Key Laboratory of Quantum Information, Department of Physics, University of
Science and Technology of China, Hefei 230026, People's Republic of China}
\author{Ping-Xing Chen}
\affiliation{Department of Applied Physics, National University of Defense Technology,
Changsha 410073, People's Republic of China}
\author{Yong-Sheng Zhang}
\email{yshzhang@ustc.edu.cn}
\affiliation{Key Laboratory of Quantum Information, Department of Physics, University of
Science and Technology of China, Hefei 230026, People's Republic of China}
\author{Zheng-Wei Zhou}
\affiliation{Key Laboratory of Quantum Information, Department of Physics, University of
Science and Technology of China, Hefei 230026, People's Republic of China}
\author{Guang-Can Guo}
\email{gcguo@ustc.edu.cn}
\affiliation{Key Laboratory of Quantum Information, Department of Physics, University of
Science and Technology of China, Hefei 230026, People's Republic of China}

\begin{abstract}
We investigate the possibility of distinguishing a set of mutually orthogonal
multipartite quantum states by local operations and classical communication
(LOCC). We connect this problem with generators of SU(N) and present a new
condition that is necessary for a set of orthogonal states to be locally
distinguishable. We show that even in multipartite cases there exists a
systematic way to check whether the presented condition is satisfied for a
given set of orthogonal states. Based on the proposed checking method, we find
that LOCC cannot distinguish three mutually orthogonal states in which two of
them are GHZ-like states.

\end{abstract}

\maketitle

Orthogonal multipartite quantum states can always be distinguished when global
measurements can be implemented, but it's a different case if only local
operations and classical communication (LOCC) are allowed. Bennett \textit{et
al}. presented a set of nine orthogonal product states that cannot be
distinguished by LOCC, which demonstrates that there is nonlocality different
from quantum entanglement \cite{bennett}. Since then the possibility of
distinguishing a set of orthogonal states by LOCC has attracted much interest
\cite{nathanson}. The problem has now been connected to many other research
fields in quantum information science such as the bound entanglement
\cite{bennett2} and the global robustness of entanglement \cite{hayashi}.

The scenario of locally distinguishing a set of orthogonal states can be
conceived as follows. Separate observers hold a multipartite quantum system,
which is prepared in one of a set of mutually orthogonal states. They know the
precise form of each state in the set but they do not know which state the
system is in. Their task is to find it out by using LOCC. First, one observer
makes a local measurement on his partite and communicates the measurement
result to others. And then another observer makes another local measurement
and so on. To say that a set of orthogonal states are locally distinguishable,
it usually needs to design a finite sequence of measurements that can always
discover which state the system is in.

Some things are known on locally distinguishing a set of orthogonal states. It
is always possible to distinguish two orthogonal multipartite quantum states
with LOCC \cite{walgate1} while it is impossible for three Bell states
\cite{ghosh1} or an unextendible product basis (UPB) \cite{bennett2}. The
necessary and sufficient conditions are obtained for a set of orthogonal
$2\times2$ quantum states that are locally distinguishable \cite{walgate2}.
There exist subspaces that have no locally distinguishable orthonormal bases
\cite{watrous}. Some necessary conditions of local distinguishability are also
found \cite{ghosh1,horodecki,chen,hayashi}. One class of necessary conditions
are connected to the characteristics that LOCC cannot increase quantum
entanglement \cite{ghosh1,ghosh2,horodecki,fan}. Recently it has been found
that if a set of orthogonal states can be distinguished by LOCC, the sum of a
distancelike entanglement measure of each state in the set cannot be larger
than the total dimension of the system \cite{hayashi}.

The known necessary conditions \cite{horodecki,hayashi} of local
distinguishability of quantum states are correlated to measures of quantum
entanglement. However how to measure quantum entanglement in multipartite
states is not well understood, so when we are given a set of orthogonal
multipartite states it is usually not easy to check whether these necessary
conditions are satisfied. In this paper we will present a result which has no
relation with entanglement measures. We first make two definitions.

Definition 1: \textit{Trivial measurement}. Measurements in quantum mechanics
are described by a collection of measurement operators $\left\{
M_{m}\right\}  $. The index $m$ refers to the outcomes that may occur in the
experiment.\ The probability that result $m$ occurs is $p\left(  m\right)
=\left\langle \Psi\right\vert M_{m}^{\dag}M_{m}\left\vert \Psi\right\rangle $,
where $\left\vert \Psi\right\rangle $ is the state of the quantum system just
before the measurement,\ and the state of the system after the measurement is
$M_{m}\left\vert \Psi\right\rangle /\sqrt{p\left(  m\right)  }$. The objects
$M_{m}^{\dag}M_{m}$ are called positive operator-valued measure (POVM)
elements, which satisfy the completeness equation $\sum_{m}M_{m}^{\dag}%
M_{m}=I$. If all POVM elements are proportional to the identity operator, we
say the measurement is trivial because it yields no information about the
state \cite{walgate2}.

Definition 2: \textit{Orthogonality-preserving measurement}. Suppose there is
a set of mutually orthogonal states $\left\{  \left\vert \varphi
_{i}\right\rangle \right\}  $ and a measurement described by $\left\{
M_{m}\right\}  $. If the states in the set $\left\{  M_{m}\left\vert
\varphi_{i}\right\rangle \right\}  $ (without normalization) with different
index $i$ are mutually orthogonal for any fixed $m$, we say the measurement
represented by $\left\{  M_{m}\right\}  $ is orthogonality-preserving to the
set $\left\{  \left\vert \varphi_{i}\right\rangle \right\}  $.

Usually there are many rounds of communications and local measurements in
distinguishing a set of orthogonal states. Local measurements in a successful
distinguishing process should be orthogonality-preserving because
orthogonality is a basic requirement for a set of states to be
distinguishable. Therefore, if a set of orthogonal multipartite states
$\left\{  \left\vert \varphi_{i}\right\rangle \right\}  $ can be distinguished
with LOCC, there should exist at least one observer who can make a local
measurement that is nontrivial and orthogonality-preserving (NTOP) to the set
$\left\{  \left\vert \varphi_{i}\right\rangle \right\}  $. If such an observer
cannot be found, the set of states cannot be distinguished using LOCC. The
similar idea has already been employed in \cite{walgate2,groisman} to give
simple proofs that LOCC cannot distinguish the set of nine product states
presented by Bennett \textit{et al}. \cite{bennett}. The question is how to
check whether such an observer exists for a general set of orthogonal states.
We have the following result.

\begin{theorem}
Suppose the $d_{a}$-dimensional particle $a$ held by Alice is a part of a
multipartite quantum system which is prepared in one of a set of mutually
orthogonal states $\left\{  \left\vert \varphi_{i}\right\rangle \right\}  $.
Alice can make a local measurement on particle $a$, which is NTOP to the set
$\left\{  \left\vert \varphi_{i}\right\rangle \right\}  $, if and only if
$t<d_{a}^{2}-1$. The integer $t$ is the maximal number of linear independent
operators in the set $\left\{  \Gamma_{ij},\Delta_{ij}\right\}  _{i\neq j}$
and operators $\Gamma_{ij}$ and $\Delta_{ij}$ are defined as follows:
\begin{align}
\Gamma_{mn}  &  =Tr_{\bar{a}}\left(  \left\vert \varphi_{m}\right\rangle
\left\langle \varphi_{n}\right\vert +\left\vert \varphi_{n}\right\rangle
\left\langle \varphi_{m}\right\vert \right)  ,\label{xx1}\\
\Delta_{mn}  &  =Tr_{\bar{a}}\left(  i\left\vert \varphi_{m}\right\rangle
\left\langle \varphi_{n}\right\vert -i\left\vert \varphi_{n}\right\rangle
\left\langle \varphi_{m}\right\vert \right)  , \label{xx2}%
\end{align}
where $Tr_{\bar{a}}$ denotes tracing on all particles except $a$.
\end{theorem}

\begin{proof}
It is not hard to find out that the operators $\Gamma_{mn}$ and $\Delta_{mn}$
are traceless Hermitian operators, so there is $t\leq d_{a}^{2}-1$. If Alice
can perform a measurement on particle $a$ and the measurement is
orthogonality-preserving to the set $\left\{  \left\vert \varphi
_{i}\right\rangle \right\}  $, there should be an operator $A^{\dag}A$, a
representative of POVM elements of her measurement, acting on the Hilbert
space of particle $a$ and satisfying%
\begin{equation}
\left\langle \varphi_{i}\right\vert A^{\dag}A\left\vert \varphi_{j}%
\right\rangle =0,~i\neq j. \label{e3}%
\end{equation}
The condition (\ref{e3}) can be rewritten as
\begin{equation}
Tr_{a}\left(  A^{\dag}A\Gamma_{ij}\right)  =Tr_{a}\left(  A^{\dag}A\Delta
_{ij}\right)  =0,~i\neq j, \label{e5}%
\end{equation}
which means the operator $A^{\dag}A$ is orthogonal to any element of the set
$\left\{  \Gamma_{ij},\Delta_{ij}\right\}  _{i\neq j}$ in the sense of
Hilbert-Schmidt inner product. Generally speaking, elements in the set
$\left\{  \Gamma_{ij},\Delta_{ij}\right\}  _{i\neq j}$ are not linear
independent and some of them may be zero. Using Gram-Schmidt orthogonalization
\cite{gram}, from the set $\left\{  \Gamma_{ij},\Delta_{ij}\right\}  _{i\neq
j}$ we can obtain a new set of operators $\left\{  \Lambda_{m}\right\}
_{m=1}^{t}$ with its elements satisfying%
\begin{equation}
\Lambda_{m}^{\dag}=\Lambda_{m},~Tr\left(  \Lambda_{m}\right)  =0,\text{ and
}Tr\left(  \Lambda_{m}\Lambda_{n}\right)  =2\delta_{mn}.
\end{equation}
The Gram-Schmidt orthogonalization ensures that any element in the set
$\left\{  \Gamma_{ij},\Delta_{ij}\right\}  _{i\neq j}$ can be expressed in a
linear composition of the elements from the set $\left\{  \Lambda_{m}\right\}
_{m=1}^{t}$ and vice versa, so the condition (\ref{e3}) or (\ref{e5}) is
equivalent to%
\begin{equation}
Tr_{a}\left(  A^{\dag}A\Lambda_{m}\right)  =0,\text{{}}m=1,\ldots,t.
\label{e7}%
\end{equation}
When $t=d_{a}^{2}-1$, the set $\left\{  \Lambda_{m}\right\}  _{m=1}^{t}$ forms
a complete set of orthogonal generators of $SU\left(  d_{a}\right)  $ and the
operator $A^{\dag}A$ satisfying Eq. (\ref{e7}) must be proportional to the
identity operator \cite{kimura}, i.e., Alice's orthogonality-preserving must
be trivial. Thus if Alice can make a NTOP measurement to the set $\left\{
\left\vert \varphi_{i}\right\rangle \right\}  $ there should be $t<d_{a}%
^{2}-1$. When $t<d_{a}^{2}-1$, we can always find a set of operators $\left\{
\lambda_{n}\right\}  _{n=1}^{r}$ with
\begin{equation}
r=d_{a}^{2}-1-t \label{r}%
\end{equation}
and
\begin{align}
(i)\text{ }\lambda_{m}^{\dag}  &  =\lambda_{m},\text{ }(ii)\text{ }Tr\left(
\lambda_{m}\right)  =0,\label{e8}\\
\text{ }(iii)\text{ }Tr\left(  \lambda_{m}\lambda_{n}\right)   &
=2\delta_{mn},\text{ }(iv)\text{ }Tr\left(  \Lambda_{m}\lambda_{n}\right)  =0.
\end{align}
The set $\left\{  \lambda_{n}\right\}  _{n=1}^{r}$ combined with $\left\{
\Lambda_{m}\right\}  _{m=1}^{t}$ form a complete set of orthogonal generators
of $SU\left(  d_{a}\right)  $. The condition (\ref{e7}) indicates the operator
$A^{\dag}A$ should be in the form%
\begin{equation}
A^{\dag}A=p\left(  \frac{I_{d_{a}}}{d_{a}}+\frac{1}{2}\sum_{n=1}^{r}%
b_{n}\lambda_{n}\right)  , \label{e10}%
\end{equation}
where $p$ is the trace of $A^{\dag}A$ and the coefficients $b_{n}$ are real.
Now Alice can perform a measurement that is NTOP to the set $\left\{
\left\vert \varphi_{i}\right\rangle \right\}  $ and she has many choices. For
example she can make a measurement that has only two POVM elements:%
\begin{align}
A_{1}^{\dag}A_{1}  &  =\frac{d_{a}}{2}\left(  \frac{I_{d_{a}}}{d_{a}}+\frac
{1}{2}\sum_{n=1}^{r}c_{n}\lambda_{n}\right)  ,\label{e9}\\
A_{2}^{\dag}A_{2}  &  =\frac{d_{a}}{2}\left(  \frac{I_{d_{a}}}{d_{a}}-\frac
{1}{2}\sum_{n=1}^{r}c_{n}\lambda_{n}\right)  ,\nonumber
\end{align}
where the coefficients $c_{n}$ are real and satisfy the condition
\begin{equation}
0<\sum_{n=1}^{r}c_{n}^{2}\leq2/\left(  d_{a}^{2}-d_{a}\right)  ,
\end{equation}
which make $A_{1}^{\dag}A_{1}$ and $A_{2}^{\dag}A_{2}$ Hermitian and positive
\cite{kimura}. It is obvious that this measurement is NTOP to the set
$\left\{  \left\vert \varphi_{i}\right\rangle \right\}  $.
\end{proof}

In the above theorem we have presented a way to check whether Alice can
perform a NTOP measurement to the set $\left\{  \left\vert \varphi
_{i}\right\rangle \right\}  $. The same checking procedure can be performed
for other observers. If no observer can make a NTOP measurement to the set
$\left\{  \left\vert \varphi_{i}\right\rangle \right\}  $, LOCC cannot
distinguish the set. Our result has no relation with entanglement measures, so
it can be applied in multipartite cases.

It is a further question that whether the rank of each POVM element of Alice's
NTOP measurement can be one when $t<d_{a}^{2}-1$. This question is especially
interesting in distinguishing orthogonal bipartite states, in which after one
observer makes such a local measurement, the possible states of the system
will be orthogonal on the other observer's side and a proper projective
measurement will reveal which state the system was originally in. That the
rank is one means the coefficients $b_{n}$ in the POVM element's expression
(\ref{e10}) satisfy%
\begin{equation}
\left(  \frac{I_{d_{a}}}{d_{a}}+\frac{1}{2}\sum_{n=1}^{r}b_{n}\lambda
_{n}\right)  ^{2}=\frac{I_{d_{a}}}{d_{a}}+\frac{1}{2}\sum_{n=1}^{r}%
b_{n}\lambda_{n}. \label{e11}%
\end{equation}
The quantity on the right hand side of Eq. (\ref{e11}) can be regarded as the
density operator of a quantum state in the Hilbert space of particle $a$. From
the solutions of Eq. (\ref{e11}), we can obtain a set of density operators of
pure states. If the POVM elements constructed from these density operators
satisfy the completeness equation, Alice can do a NTOP measurement to the set
$\left\{  \left\vert \varphi_{i}\right\rangle \right\}  $ with all the ranks
of POVM elements being one. When particle $a$ held by Alice is a qubit, i.e.,
$d_{a}=2$, Eq. (\ref{e11}) is equivalent to $\sum_{n=1}^{r}b_{n}^{2}=1$. We
have the following result:

\begin{corollary}
Alice and Bob share a $2\times n$ quantum system which is prepared in one of a
set of mutually orthogonal states $\left\{  \left\vert \varphi_{i}%
\right\rangle \right\}  $. They can discover which state the system is in
using one-way classical communication and local projective measurements if
Alice can perform a NTOP measurement to the set $\left\{  \left\vert
\varphi_{i}\right\rangle \right\}  $ (i.e., if $r\geq1$).
\end{corollary}

\begin{proof}
That Alice can perform a NTOP measurement to the set $\left\{  \left\vert
\varphi_{i}\right\rangle \right\}  $ means the integer $r$ defined in
Eq.\ (\ref{r}) is equal or larger than one, so Alice can perform a measurement
described by the following two POVM elements
\begin{align}
A_{1}^{\dag}A_{1}  &  =\frac{I_{2}}{2}+\frac{1}{2}\sum_{n=1}^{r}b_{n}%
\lambda_{n},\label{e12}\\
A_{2}^{\dag}A_{2}  &  =\frac{I_{2}}{2}-\frac{1}{2}\sum_{n=1}^{r}b_{n}%
\lambda_{n}, \label{e121}%
\end{align}
on her qubit $a$, where $\sum_{n=1}^{r}b_{n}^{2}=1$. This measurement is NTOP
to the set $\left\{  \left\vert \varphi_{i}\right\rangle \right\}  $ and the
rank of each POVM element is one. Actually Alice's measurement is a standard
projective measurement. After Alice has performed her measurement, the
possible states of the system will be orthogonal on Bob's side. According to
the result in Alice's measurement, Bob can perform a suitable projective
measurement on his particle to find out which state the system was originally in.
\end{proof}

The corollary is actually equivalent to \textit{theorem 1} in \cite{walgate2},
but we have given a systematic method to check whether $r\geq1$. In addition
we have given an explicit expression (\ref{e12},\ref{e121}) for Alice's
measurement when she can go first.

Till now we have only considered the first local measurement in the whole
distinguishing process. When one local observer can perform a NTOP measurement
to the set $\left\{  \left\vert \varphi_{i}\right\rangle \right\}  $, it means
that the states are possible to be distinguished using LOCC, so it is
necessary to give a consideration on the second local measurement. We use
operators $\left\{  M_{m}\right\}  $ to describe the first local measurement
which is NTOP to the set $\left\{  \left\vert \varphi_{i}\right\rangle
\right\}  $. When result $m$ occurs in the first measurement, the observers
are forced to distinguish a new set of orthogonal states $\left\{
M_{m}\left\vert \varphi_{i}\right\rangle \right\}  $ (without normalization).
We have mentioned that in a successful distinguishing process each local
measurement should be orthogonality-preserving. As we discuss the first
measurement we can check whether there exists an observer who can implement a
NTOP measurement to the set $\left\{  M_{m}\left\vert \varphi_{i}\right\rangle
\right\}  $. The detailed checking method is described in the above theorem
and its proof. If there is no such an observer except the first measurement
performer, we can say the first measurement represented by the operators
$\left\{  M_{m}\right\}  $ is inappropriate. Following this idea, we find that
two orthogonal GHZ-like states and an arbitrary state from their complementary
subspace cannot be distinguished by LOCC, i.e., the following three mutually
orthogonal states cannot be distinguished with LOCC:%
\begin{align}
\left\vert \varphi_{1}\right\rangle  &  =s\left\vert 0\right\rangle
_{a}\left\vert 0\right\rangle _{b}\left\vert 0\right\rangle _{c}+t\left\vert
1\right\rangle _{a}\left\vert 1\right\rangle _{b}\left\vert 1\right\rangle
_{c},\\
\left\vert \varphi_{2}\right\rangle  &  =t^{\ast}\left\vert 0\right\rangle
_{a}\left\vert 0\right\rangle _{b}\left\vert 0\right\rangle _{c}-s^{\ast
}\left\vert 1\right\rangle _{a}\left\vert 1\right\rangle _{b}\left\vert
1\right\rangle _{c},\\
\left\vert \varphi_{3}\right\rangle  &  =x_{1}\left\vert 1\right\rangle
_{a}\left\vert 0\right\rangle _{b}\left\vert 0\right\rangle _{c}%
+x_{2}\left\vert 0\right\rangle _{a}\left\vert 1\right\rangle _{b}\left\vert
1\right\rangle _{c}+x_{3}\left\vert 0\right\rangle _{a}\left\vert
1\right\rangle _{b}\left\vert 0\right\rangle _{c}\\
&  +x_{4}\left\vert 1\right\rangle _{a}\left\vert 0\right\rangle
_{b}\left\vert 1\right\rangle _{c}+x_{5}\left\vert 0\right\rangle
_{a}\left\vert 0\right\rangle _{b}\left\vert 1\right\rangle _{c}%
+x_{6}\left\vert 1\right\rangle _{a}\left\vert 1\right\rangle _{b}\left\vert
0\right\rangle _{c},\nonumber
\end{align}
where $s\times t\neq0$, $\left\vert s\right\vert ^{2}+\left\vert t\right\vert
^{2}=1$, and $\sum_{i=1}^{6}\left\vert x_{i}\right\vert ^{2}=1$.

The proof of the local indistinguishability of the above states $\left\{
\left\vert \varphi_{i}\right\rangle \right\}  _{i=1}^{3}$ can be based on
theorem 1. Suppose the qubits $a$, $b$, and $c$ are held by Alice, Bob and
Charlie, respectively. Without loss of generality we can assume $s$ and $t$
are real \cite{explain}. We now check whether Alice can go first by utilizing
theorem 1. We first define a set of operators for qubit $a$,
\begin{equation}
S_{A}=\{\Gamma_{12},\Delta_{12},\Gamma_{13},\Delta_{13},\Gamma_{23}%
,\Delta_{23}\},
\end{equation}
where $\Gamma_{ij}$ and $\Delta_{ij}$ are calculated from $\left\{  \left\vert
\varphi_{i}\right\rangle \right\}  _{i=1}^{3}$ according to Eq. (\ref{xx1})
and Eq. (\ref{xx2}) in theorem 1. We need to figure out the number of linear
independent operators in $S_{A}$ which we denote by $t_{A}$. For our purpose
the last four operators in $S_{A}$ can be replaced by
\begin{align}
s\times\Gamma_{13}+t\times\Gamma_{23} &  =\operatorname{Re}\left(
x_{1}\right)  \sigma_{x}^{a}+\operatorname{Im}\left(  x_{1}\right)  \sigma
_{y}^{a},\\
s\times\Delta_{13}+t\times\Delta_{23} &  =\operatorname{Im}\left(
x_{1}\right)  \sigma_{x}^{a}-\operatorname{Re}\left(  x_{1}\right)  \sigma
_{y}^{a},\\
t\times\Gamma_{13}-s\times\Gamma_{23} &  =\operatorname{Re}\left(
x_{2}\right)  \sigma_{x}^{a}-\operatorname{Im}\left(  x_{2}\right)  \sigma
_{y}^{a},\\
t\times\Delta_{13}-s\times\Delta_{23} &  =\operatorname{Im}\left(
x_{2}\right)  \sigma_{x}^{a}+\operatorname{Re}\left(  x_{2}\right)  \sigma
_{y}^{a}.
\end{align}
Notice that the first two operators in $S_{A}$ are $\Gamma_{12}=2st\sigma
_{z}^{a}$ and $\Delta_{12}=0$, it is not hard to find that $t_{A}=3$ when
$x_{1}\neq0$ or $x_{2}\neq0$ and $t_{A}=1$ when $x_{1}=x_{2}=0$. According to
theorem 1, we can conclude that Alice cannot go first when $x_{1}\neq0$ or
$x_{2}\neq0$.

When $x_{1}=x_{2}=0$, Alice can go first. We use operators $\left\{
A_{m}\right\}  $ to describe Alice's NTOP measurement, then $A_{m}^{\dag}%
A_{m}$ should be in the form%
\begin{equation}
A_{m}^{\dag}A_{m}=p_{m}I_{2}^{a}+b_{mx}\sigma_{x}^{a}+b_{my}\sigma_{y}^{a},
\end{equation}
where $p_{m}>0,$ $b_{mx}$ and $b_{my}$ are real. If result $m$ occurs in
Alice's measurement, they will be forced to distinguish the orthogonal states
$\left\{  A_{m}\left\vert \varphi_{i}\right\rangle \right\}  _{i=1}^{3}$. As
we check whether Alice can perform a NTOP measurement to $\left\{  \left\vert
\varphi_{i}\right\rangle \right\}  _{i=1}^{3}$ by utilizing theorem 1, we can
check whether Bob or Charlie can do a NTOP measurement to $\left\{
A_{m}\left\vert \varphi_{i}\right\rangle \right\}  _{i=1}^{3}$. The result is
that only when%

\begin{equation}
x_{3}p_{m}+x_{6}\left\langle 0\right\vert _{a}A_{m}^{\dag}A_{m}\left\vert
1\right\rangle _{a}=x_{4}p_{m}+x_{5}\left\langle 1\right\vert _{a}A_{m}^{\dag
}A_{m}\left\vert 0\right\rangle _{a}=0, \label{e20}%
\end{equation}
Bob can perform a NTOP measurement to $\left\{  A_{m}\left\vert \varphi
_{i}\right\rangle \right\}  _{i=1}^{3}$ and only when%
\begin{equation}
x_{3}\left\langle 1\right\vert _{a}A_{m}^{\dag}A_{m}\left\vert 0\right\rangle
_{a}+x_{6}p_{m}=x_{4}\left\langle 0\right\vert _{a}A_{m}^{\dag}A_{m}\left\vert
1\right\rangle _{a}+x_{5}p_{m}=0, \label{e21}%
\end{equation}
Charlie can do a NTOP measurement to $\left\{  A_{m}\left\vert \varphi
_{i}\right\rangle \right\}  _{i=1}^{3}$. Notice that $\sum_{i=3}^{6}\left\vert
x_{i}\right\vert ^{2}=1$, without loss of generality we assume $x_{3}\neq0$.
(i) We first consider the case where $x_{6}=0$. Since Alice's measurement is
nontrivial, there is $\left\langle 1\right\vert _{a}A_{m}^{\dag}%
A_{m}\left\vert 0\right\rangle _{a}\neq0$ at least for one $m$. Neither Eq.
(\ref{e20}) nor (\ref{e21}) can be satisfied for this special $m$, so when
this special result $m$ occurs in Alice's measurement neither Bob nor Charlie
can do the second NTOP measurement; LOCC cannot distinguish the original
states $\left\{  \left\vert \varphi_{i}\right\rangle \right\}  _{i=1}^{3}$.
(ii) When $x_{6}\neq0$, from Eq. (\ref{e20}) we can obtain
\begin{equation}
\left\langle 0\right\vert _{a}A_{m}^{\dag}A_{m}\left\vert 1\right\rangle
_{a}=-\frac{x_{3}}{x_{6}}\times p_{m}, \label{e28}%
\end{equation}
and from Eq. (\ref{e21}) we can obtain
\begin{equation}
\left\langle 0\right\vert _{a}A_{m}^{\dag}A_{m}\left\vert 1\right\rangle
_{a}=-\frac{x_{6}^{\ast}}{x_{3}^{\ast}}\times p_{m}. \label{e29}%
\end{equation}
If we assume LOCC can distinguish the states $\left\{  \left\vert \varphi
_{i}\right\rangle \right\}  _{i=1}^{3}$, for any $m$ either Eq. (\ref{e28}) or
(\ref{e29}) should be satisfied. Notice that all quantities $\left\langle
0\right\vert _{a}A_{m}^{\dag}A_{m}\left\vert 1\right\rangle _{a}$ with
different $m$ obtained from Eq. (\ref{e28}) or (\ref{e29}) have the same phase
factor, there is
\begin{equation}
\sum_{m}\left\langle 0\right\vert _{a}A_{m}^{\dag}A_{m}\left\vert
1\right\rangle _{a}\neq0, \label{e30}%
\end{equation}
which contradicts with the completeness equation $\sum_{m}A_{m}^{\dag}%
A_{m}=I_{2}^{a}$. So LOCC cannot distinguish the states $\left\{  \left\vert
\varphi_{i}\right\rangle \right\}  _{i=1}^{3}$.

We have shown that even when Alice can go first LOCC still cannot distinguish
the states $\left\{  \left\vert \varphi_{i}\right\rangle \right\}  _{i=1}^{3}%
$. Due to the symmetry of $\left\vert \varphi_{1}\right\rangle $ and
$\left\vert \varphi_{2}\right\rangle $ and the arbitrariness of $\left\vert
\varphi_{3}\right\rangle $, we can end our proof.

In the above we have utilized theorem 1 to prove the local
indistinguishability of a three-state example. To show the usefulness and
applicability of the theorem we will discuss two more examples. Bennett
\textit{et al.} mentioned an example involving four states in the discussion
part of their original paper \cite{bennett}%
\begin{align}
\left\vert \varphi_{1}\right\rangle  &  =\left\vert 0\right\rangle
_{a}\left\vert 1\right\rangle _{b}\left\vert +\right\rangle _{c},\left\vert
\varphi_{2}\right\rangle =\left\vert 1\right\rangle _{a}\left\vert
+\right\rangle _{b}\left\vert 0\right\rangle _{c},\label{x1}\\
\left\vert \varphi_{3}\right\rangle  &  =\left\vert +\right\rangle
_{a}\left\vert 0\right\rangle _{b}\left\vert 1\right\rangle _{c},\left\vert
\varphi_{4}\right\rangle =\left\vert -\right\rangle _{a}\left\vert
-\right\rangle _{b}\left\vert -\right\rangle _{c},
\end{align}
where $\left\vert +\right\rangle =\left(  \left\vert 0\right\rangle
+\left\vert 1\right\rangle \right)  /\sqrt{2}$, $\left\vert -\right\rangle
=\left(  \left\vert 0\right\rangle -\left\vert 1\right\rangle \right)
/\sqrt{2}$. At that time they suspect that these four states cannot be
distinguished by LOCC but they cannot give a proof. However, by directly
utilizing our theorem it can be found that there doesnot exist a local
observer that can make a NTOP measurement to these four states. We take the
holder of qubit $a$ as an example. According to theorem 1, we need to figure
out the number of linear independent operator in the set%
\begin{equation}
\{\Gamma_{12}^{a},\Delta_{12}^{a},\Gamma_{13}^{a},\Delta_{13}^{a},\Gamma
_{14}^{a},\Delta_{14}^{a},\Gamma_{23}^{a},\Delta_{23}^{a},\Gamma_{24}%
^{a},\Delta_{24}^{a},\Gamma_{34}^{a},\Delta_{34}^{a}\}, \label{x2}%
\end{equation}
where the definitions of $\Gamma_{ij}^{a}$ and $\Delta_{ij}^{a}$ are in our
theorem, and the superscript $a$ means they are operators acting on the
Hilbert space of qubit $a$. Simple calculations show%
\begin{align}
\Gamma_{12}^{a}  &  =Tr_{bc}\left(  \left\vert \varphi_{1}\right\rangle
\left\langle \varphi_{2}\right\vert +\left\vert \varphi_{2}\right\rangle
\left\langle \varphi_{1}\right\vert \right)  =\sigma_{x}^{a}/2,\\
\Delta_{12}^{a}  &  =iTr_{bc}\left(  \left\vert \varphi_{1}\right\rangle
\left\langle \varphi_{2}\right\vert -\left\vert \varphi_{2}\right\rangle
\left\langle \varphi_{1}\right\vert \right)  =-\sigma_{y}^{a}/2,\\
\Gamma_{34}^{a}  &  =Tr_{bc}\left(  \left\vert \varphi_{3}\right\rangle
\left\langle \varphi_{4}\right\vert +\left\vert \varphi_{4}\right\rangle
\left\langle \varphi_{3}\right\vert \right)  =-\sigma_{z}^{a}/2,
\end{align}
so the number of linear independent operators in Eq. (\ref{x2}) is three.
According to theorem 1, the holder of qubit $a$ cannot make a NTOP
measurement. Since no observer can make a NTOP measurement to the states in
Eq. (\ref{x1}), these states cannot be distinguished with LOCC.

Bennett \textit{et al.} proved a general result that an unextendible product
basis cannot be distinguished by LOCC in a later paper \cite{bennett2}. Since
the four states in Eq. (\ref{x1}) form an unextendible product basis, their
local indistinguishability can also be proved by utilizing the result in
\cite{bennett2}. However this fact doesnot decrease the usefulness of our
theorem. First, the result in \cite{bennett2} is quite different from our
theorem. Second, the result in \cite{bennett2} cannot be utilized to prove the
local indistinguishability of a variation of the four states in Eq.
(\ref{x1}). For an instance, let us examine the following four states%
\begin{align}
\left\vert \bar{\varphi}_{1}\right\rangle  &  =\left\vert \varphi
_{1}\right\rangle ,\left\vert \bar{\varphi}_{2}\right\rangle =\sqrt
{2}\left\vert \varphi_{2}\right\rangle +\left\vert \varphi_{4}\right\rangle
,\label{x3}\\
\left\vert \bar{\varphi}_{3}\right\rangle  &  =\left\vert \varphi
_{3}\right\rangle ,\left\vert \bar{\varphi}_{4}\right\rangle =\left\vert
\varphi_{2}\right\rangle -\sqrt{2}\left\vert \varphi_{4}\right\rangle ,
\label{xx3}%
\end{align}
where $\left\vert \varphi_{j}\right\rangle $, $j=1,2,3,4$, are states in Eq.
(\ref{x1}). Since $\left\vert \bar{\varphi}_{2}\right\rangle $ and $\left\vert
\bar{\varphi}_{4}\right\rangle $ are entangled states, the states in Eq.
(\ref{x3}) and (\ref{xx3}) are not an unextendible product basis; So we cannot
utilize the result in \cite{bennett2} to prove the local indistinguishability
of these states. However we can still use our theorem to prove the local
indistinguishability of these states. For qubit $a$, there are%
\begin{align}
\Gamma_{12}^{a}  &  =Tr_{bc}\left(  \left\vert \bar{\varphi}_{1}\right\rangle
\left\langle \bar{\varphi}_{2}\right\vert +\left\vert \bar{\varphi}%
_{2}\right\rangle \left\langle \bar{\varphi}_{1}\right\vert \right)
=\sigma_{x}^{a}/\sqrt{2},\\
\Delta_{12}^{a}  &  =iTr_{bc}\left(  \left\vert \bar{\varphi}_{1}\right\rangle
\left\langle \bar{\varphi}_{2}\right\vert -\left\vert \bar{\varphi}%
_{2}\right\rangle \left\langle \bar{\varphi}_{1}\right\vert \right)
=-\sigma_{y}^{a}/\sqrt{2},\\
\Gamma_{34}^{a}  &  =Tr_{bc}\left(  \left\vert \bar{\varphi}_{3}\right\rangle
\left\langle \bar{\varphi}_{4}\right\vert +\left\vert \bar{\varphi}%
_{4}\right\rangle \left\langle \bar{\varphi}_{3}\right\vert \right)
=\sigma_{z}^{a}/\sqrt{2}.
\end{align}
So the number of linear independent operators in $\left\{  \Gamma_{ij}%
^{a},\Delta_{ij}^{a}\right\}  _{i\neq j}$ is three. According to theorem 1,
the holder of qubit $a$ cannot go first. For qubit $b$, there are%
\begin{align}
\Gamma_{13}^{b}  &  =Tr_{ac}\left(  \left\vert \bar{\varphi}_{1}\right\rangle
\left\langle \bar{\varphi}_{3}\right\vert +\left\vert \bar{\varphi}%
_{3}\right\rangle \left\langle \bar{\varphi}_{1}\right\vert \right)
=\sigma_{x}^{b}/2,\\
\Delta_{13}^{b}  &  =iTr_{ac}\left(  \left\vert \bar{\varphi}_{1}\right\rangle
\left\langle \bar{\varphi}_{3}\right\vert -\left\vert \bar{\varphi}%
_{3}\right\rangle \left\langle \bar{\varphi}_{1}\right\vert \right)
=\sigma_{y}^{b}/2,\\
\Gamma_{24}^{b}  &  =Tr_{ac}\left(  \left\vert \bar{\varphi}_{2}\right\rangle
\left\langle \bar{\varphi}_{4}\right\vert +\left\vert \bar{\varphi}%
_{4}\right\rangle \left\langle \bar{\varphi}_{2}\right\vert \right)
=2\sqrt{2}\sigma_{x}^{b}+\sigma_{z}^{b}/2,
\end{align}
so we can also conclude that the holder of qubit $b$ cannot go first according
to theorem 1. For qubit $c$, there are%
\begin{align}
\Gamma_{12}^{c}  &  =Tr_{ab}\left(  \left\vert \bar{\varphi}_{1}\right\rangle
\left\langle \bar{\varphi}_{2}\right\vert +\left\vert \bar{\varphi}%
_{2}\right\rangle \left\langle \bar{\varphi}_{1}\right\vert \right)
=-\sigma_{z}^{c}/2,\\
\Delta_{12}^{c}  &  =iTr_{ab}\left(  \left\vert \bar{\varphi}_{1}\right\rangle
\left\langle \bar{\varphi}_{2}\right\vert -\left\vert \bar{\varphi}%
_{2}\right\rangle \left\langle \bar{\varphi}_{1}\right\vert \right)
=-\sigma_{y}^{c}/2,\\
\Gamma_{23}^{c}  &  =Tr_{ab}\left(  \left\vert \bar{\varphi}_{2}\right\rangle
\left\langle \bar{\varphi}_{3}\right\vert +\left\vert \bar{\varphi}%
_{3}\right\rangle \left\langle \bar{\varphi}_{2}\right\vert \right)
=\sigma_{x}^{c}/\sqrt{2},
\end{align}
So the number of linear independent operators in $\left\{  \Gamma_{ij}%
^{c},\Delta_{ij}^{c}\right\}  _{i\neq j}$ is three; The Holder of qubit $c$
cannot go first. Since no local observer can go first, the orthogonal states
in Eq. (\ref{x3}) are local indistinguishable.

In conclusion the question that whether a given set of orthogonal states can
be discriminated by LOCC is of considerable importance but it has not be
solved for general cases. Usually there are many rounds of communications and
local measurements in a distinguishing process. In a successful distinguishing
process local measurements are required to be orthogonality-preserving. We
have presented a systematic method to check whether an observer can make a
local measurement that is NTOP to a set of orthogonal states. If no observer
can make a NTOP measurement, the set of orthogonal states cannot be
distinguished by LOCC. Our checking method has no relation with entanglement
measures, so it is effective in multipartite cases.

This work was funded by the National Fundamental Research Program
(2006CB921900), the National Natural Science Foundation of China (Grant Nos.
10674127, 60621064, 10574126 and 10404039), the Innovation funds from Chinese
Academy of Science and program for NCET, and International Cooperation Program
from CAS and Ministry of Science and Technology of China.


\begin{thebibliography}{99}
\bibitem {explain}Suppose $s=\left\vert s\right\vert e^{i\theta_{1}}$,
$t=\left\vert t\right\vert e^{i\theta_{2}}$. The local distinguishability of
$\left\{  \left\vert \varphi_{1}\right\rangle ,\left\vert \varphi
_{2}\right\rangle ,\left\vert \varphi_{3}\right\rangle \right\}  $ is the same
as that of $\left\{  \left\vert \varphi_{1}^{\prime}\right\rangle ,\left\vert
\varphi_{2}^{\prime}\right\rangle ,\left\vert \varphi_{3}^{\prime
}\right\rangle \right\}  $ with $\left\vert \varphi_{1}^{\prime}\right\rangle
=\left\vert s\right\vert \left\vert \bar{0}\right\rangle _{a}\left\vert
0\right\rangle _{b}\left\vert 0\right\rangle _{c}+\left\vert t\right\vert
\left\vert \bar{1}\right\rangle _{a}\left\vert 1\right\rangle _{b}\left\vert
1\right\rangle _{c}$, $\left\vert \varphi_{2}^{\prime}\right\rangle
=\left\vert t\right\vert \left\vert \bar{0}\right\rangle _{a}\left\vert
0\right\rangle _{b}\left\vert 0\right\rangle _{c}-\left\vert s\right\vert
\left\vert \bar{1}\right\rangle _{a}\left\vert 1\right\rangle _{b}\left\vert
1\right\rangle _{c}$, where $\left\vert \bar{0}\right\rangle _{a}=\left\vert
0\right\rangle _{a},\left\vert \bar{1}\right\rangle _{a}=e^{i\left(
\theta_{2}-\theta_{1}\right)  }\left\vert 1\right\rangle _{a}$. The state
$\left\vert \varphi_{3}^{\prime}\right\rangle $ is the same as $\left\vert
\varphi_{3}\right\rangle $ but the basis for qubit $a$ is changed from
$\left\{  \left\vert 0\right\rangle _{a},\left\vert 1\right\rangle
_{a}\right\}  $ into $\left\{  \left\vert \bar{0}\right\rangle _{a},\left\vert
\bar{1}\right\rangle _{a}\right\}  $.
\end{thebibliography}
\end{document}